\documentstyle[psfig,mynamed]{article}
\renewcommand{\thefootnote}{}     
\twocolumn
\newcommand{\citeN}[1]{\citeauthor{#1}\ (\citeyear{#1})}
\newcommand{\citeNP}[1]{\citeauthor{#1}\ \citeyear{#1}}

\newcommand{\aap}{{A\&A }}
\newcommand{\apj}{{ApJ }}
\newcommand{\araa}{{ARA\&A }}
\newcommand{\an}{{Astr.\ Nachr.\ }}
\newcommand{\jgr}{{J.\ Geophys.\ Res.\ }}
\newcommand{\mnras}{{MNRAS }}
\newcommand{\solphys}{{Sol.\ Phys.\ }}

\newcommand{\ujlisti}{
\itemsep=0 em
\parsep=0.5 em
\partopsep=0.25 em
\topsep=0 em}

\newenvironment{lista}{\begin{list}{--}{\ujlisti}}{\end{list}}

\newcommand{\szekcio}[1]{\begin{center}{
 \section*{\normalsize{\large\bf #1}}}\end{center}}
\newcommand{\alszekcio}[1]{\begin{center}{
 \subsection*{\normalsize{\large\bf #1}}}\end{center}}
\newcommand{\szekciononu}[1]{\section*{{\bf #1}}}

\newcommand{\<}{\begin{equation} }
\renewcommand{\>}{\end{equation} }

\newcommand{\vc}[1]{\mbox{\bf #1}}
\newcommand{\mx}[1]{\hat{#1}}

\newcommand{\ptl}{\partial}

\def\la{\mathrel{\mathchoice {\vcenter{\offinterlineskip\halign{\hfil
 $\displaystyle##$\hfil\cr<\cr\sim\cr}}}
 {\vcenter{\offinterlineskip\halign{\hfil$\textstyle##$\hfil\cr
 <\cr\sim\cr}}}
 {\vcenter{\offinterlineskip\halign{\hfil$\scriptstyle##$\hfil\cr
 <\cr\sim\cr}}}
 {\vcenter{\offinterlineskip\halign{\hfil$\scriptscriptstyle##$\hfil\cr
 <\cr\sim\cr}}}}}

\begin{document}

\begin{minipage}[t]{17 cm}
\strut\vskip 2 cm 
\title{ {\large\bf
\strut WHAT MAKES THE SUN TICK?\\ The Origin of the Solar Cycle} }

\author{  {\normalsize\bf 
\strut K. Petrovay}\\
{ \normalsize\rm \strut E\"otv\"os University, Department of Astronomy,
Budapest, Pf. 32, H-1518 Hungary}\\
 {\normalsize\rm \strut Phone: +36 20 3404907 \hskip 2 em Fax: +36 1 3722940 
  \hskip 2 em E-mail: kris@astro.elte.hu}
}

\date{}

\maketitle
\end{minipage}

\normalsize
\vskip -2.7 cm

\parindent=0 cm
\bf Abstract. \rm 
In contrast to the situation with the geodynamo, no breakthrough has been made
in the solar dynamo problem for decades. Since the appearance of mean-field
electrodynamics in the 1960's, the only really significant advance was in the
field of flux tube theory and flux emergence calculations. These new results,
together with helioseismic evidence, have led to the realization that the
toroidal magnetic flux giving rise to activity phenomena must be stored and
presumably generated below the convection zone proper, in what I will call the
DOT (Dynamo-Overshoot-Tachoclyne) layer. The only segment of the problem we can
claim to basically understand is the transport of flux from this layer to the
surface. On the other hand, as reliable models for the DOT layer do not exist
we are clueless concerning the precise mechanisms responsible for
toroidal/poloidal flux conversion and for characteristic migration patterns
(extended butterfly diagram) and periodicities. Even the most basic result of
mean-field theory, the identification of the butterfly diagram with an
$\alpha$--$\omega$ dynamo wave, has been questioned. This review therefore will
necessarily ask more questions than give answers. Some of these key questions
are 
\begin{lista}
\item Structure of the DOT layer
\item $\alpha$-quenching and distributed dynamo 
\item High-latitude migration patterns and their interpretation 
\item The ultimate fate of emerged flux
\end{lista}
\footnote{\noindent\small 
In: {\it The Solar Cycle and Terrestrial Climate},\\ 
\strut\hskip 4 em ESA Publ.\ SP-463, p. 3--14 (2000)}

\parindent=1 em

\renewcommand{\thefootnote}{\arabic{footnote}}     

\szekcio{1. INTRODUCTION}
The turn of the millennium invites us to look back and draw balances in all
fields of human activity. Yet in solar dynamo theory  we also have an added
incentive to make such an assessment. In the theory of the geodynamo a
significant breakthrough has been achieved in the past few years
(\citeNP{Glatzmaier+Roberts:geodynamo}, \citeNP{Kuang+Bloxham:geodynamo},
\citeNP{Olson+:geodynamo}), leading to a surge of renewed activity in the
field. One cannot but wonder if a similar breakthrough is within reach in the
case of the solar dynamo. Unfortunately, as it will turn out from this review,
the prospects are rather bleak, at least on a short term.

As for such a comparative assessment one needs a wider historical outlook this 
review will not be restricted to the developments that have taken place since
the reviews of \citeN{Weiss:CUPrev} and \citeN{Schmitt:Potsdam}. (Such
developments were mostly limited to advances in the study of interface dynamos,
cf.\ Sect.~3 below.) A wider historical overview, starting with the dawn of
mean-field theory in the 1950s and 60s will thus be given in Section 2 below.
Given the finite amount of space available, I will compensate for this wider
temporal scope of the review by restricting attention strictly to the problem
of the origin of the solar cycle, i.e.\ of the 22 year periodic variation of
solar activity, and associated migration patterns (butterfly diagram). Solar
activity variations on both shorter and longer timescales are ignored, as are
the solar-type magnetic cycles in other stars, non-axisymmetric phenomena such
as active longitudes, and the problem of the long-time phase coherence of the
cycle. This restriction is imposed out of necessity only, and in no way does it
imply that these effects do not yield  important clues even to the origin of
the 22 year cycle itself. Clearly, a critical test of any theory of the solar
cycle is whether it can be readily extended to predict these other phenomena as
well.

After the historical overview, Section 3 will attempt to cut some order in the
dazzling multitude of solar dynamo models by introducing a classification
scheme. Three main model families can be clearly discerned: overshoot dynamos,
interface dynamos and flux transport models, circulation-driven ``conveyor
belt'' models being the  most important subgroup of the latter class. Finally,
Section 4 calls attention to some key areas where more intensive theoretical or
observational efforts could lead to significant advance.

But first of all we should state clearly what are the basic observational
facts  to be interpreted by a solar cycle model.  Once we apply our
aforementioned restriction excluding long-term variations, stellar activity
etc., the remaining list is quite short. 

\begin{lista}

\item {\bf The 11/22 year cycle period.} Beside reproducing the value of the
period, the crude agreement of this value with the timescale of pole-equator
diffusion in the convective zone also asks for an explanation. While such an
order-of-magnitude equality can certainly be coincidental (cf.\ the coincidence
of the solar rotation period with the convective turnover time in the deep
convective zone), a natural explanation for it would clearly make any cycle
model more attractive.

\begin{figure}[htb]
\centerline{\psfig{figure=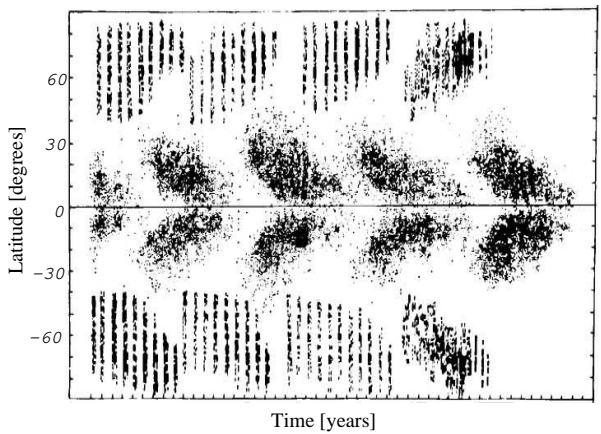,width=\columnwidth}}
\caption{\small\it
Extended butterfly diagram of solar activity: time-latitude distribution of
sunspot groups (low-latitude branches) and polar faculae (high-latitude
branches). After Makarov \& Sivaraman (1989)
\label{fig:butterfly} }
\end{figure}

\item {\bf The characteristic migration pattern (extended butterfly diagram).} 
Our knowledge of latitudinal migration patterns of magnetic flux is summarized
in the extended butterfly diagram of Figure~\ref{fig:butterfly}. The tracers
shown here track partly toroidal and partly poloidal fields.\footnote{Note
that, owing to our assumption of axial symmetry, ``toroidal'' is now synonymous
with ``zonal'' or ``azimuthal'', denoting the $\phi$-component of a vector
field in spherical coordinates, while ``poloidal'' is synonymous with
``meridional'', denoting the remaining components. This relieves us from giving
a more generic definition of these terms.} While azimuthal field lines by
definition cannot cross the surface, the observed properties of large-scale
solar active regions\footnote{To be specific: their preferential East--West 
orientation following Hale's polarity rules. (The western, or leading magnetic
polarity is identical on the same hemisphere and within the same cycle, and
alternates between hemispheres and cycles)} strongly suggest that they are
tracers of a subsurface toroidal field, locally bulging out into the
atmosphere. In this sense, photospheric magnetometry can give us information
about the migration patterns of the toroidal field component as well. At low
($\la 35^\circ$) latitudes both the poloidal and the toroidal field components
migrate equatorward. At high latitudes, poloidal fields show a marked poleward
migration, as indicated also by the migration pattern of a number of tracers
such as quiescent prominences or the coronal green line. The migration pattern
of high-latitude toroidal fields is less clear ---a point we will return to in
Section 4.2 below.

\item {\bf The confinement of strong activity (large active regions) to low
  heliographic latitudes $|\Phi|\la 35^\circ$.} 

\item {\bf The phase dilemma(s).} In its original sense
(\citeNP{Stix:phase.dilemma}) the phase dilemma consists of the fact that at
low latitudes the radial field (derived by azimuthal averaging of the
magnetograms) is in an approximately $\pi$ phase lag compared to the toroidal
field at the same latitude. Another phase lag to be explained is the $\pi/2$
lag between the two branches of the butterfly diagram, i.e. that the polar
field reversal occurs slightly after the sunspot maximum. Finally, the phase of
torsional oscillations (cycle-related periodic oscillations of the rotational
velocity in migrating belts) relative to the toroidal field is a third
quantity constraining theories of the cycle.

\end{lista}

\szekcio{2. HISTORY}
\alszekcio{2.1. Convection zone dynamos\vspace{-0.5 cm}}
It all started with Parker's (\citeyear{Parker:cyclonic}) classic paper that
set down the foundations for solar $\alpha\Omega$ dynamo theory. In its trace,
mean field electrodynamics was developed during the 1960's
(\citeNP{Steenbeck+}, \citeNP{Krause+Raedler:book}). To give a reminder of the
basics, described in so many other reviews (e.g.\ \citeNP{Cowling:ARAA},
\citeNP{Belvedere:SPhrev}) the induction equation in a turbulent conductive
medium reads 
\< \ptl_t\vc B =\nabla\times(\vc U\times\vc B +\vec{\cal E}) 
   -\nabla\times\eta\nabla\times\vc B  .   \label{eq:avind} \>
where $\vc B$ is the mean magnetic field, $\vc U$ is the large-scale flow
velocity, $\ptl_t$ denotes time derivative, $\eta$ is the magnetic diffusivity,
and $\vec{\cal E}$ is the turbulent electromotive force arising as a result of
the interaction between the turbulent velocity field $\vc v$ and the turbulent
magnetic field. This latter is in turn the result of the action of $\vc v$ on
the mean field $\vc B$, so $\vec{\cal E}$ is a functional of $\vc v$ and $\vc
B$. Assuming scale separation $l\ll H_B$ where $H_B$ is the length scale of the
mean field and $l$ is the scale of turbulence, $\vec{\cal E}$ can be expanded
in the derivatives of $\vc B$. For homogeneous and isotropic turbulence this
yields
\< \vec{\cal E}=\alpha\vc B -\beta\,\nabla\times\vc B      \label{eq:Ekif} \>
where $\alpha$ and $\beta$ are now functionals of $\vc v$ only. Substituting
(\ref{eq:Ekif}) into (\ref{eq:avind}) we see that the role of $\beta$ is
formally identical to that of $\eta$. For this reason $\beta$ is called {\it
turbulent magnetic diffusivity,} and elementary considerations or even
dimensional analysis yield $\beta\sim lv$. As for turbulence the Reynolds
number $\beta/\eta\gg 1$, in practice $\eta$ can be omitted in equation
(\ref{eq:avind}). In contrast, the pseudoscalar $\alpha$ gives rise to a
qualitatively new effect, the {\it $\alpha$-effect.} 

In the axisymmetric case considered here, using spherical coordinates $\theta$,
$\phi$, $r$, $\vc B$ can be split as
\[ \vc B=B\vc e_\phi +\nabla\times(A\vc e_\phi)       \]
where $\vc e_\phi$ is the azimuthal unit vector, $B$ is the toroidal field
component, and $A$ is the (toroidal) vector potential of the poloidal field. 
We further assume that $\vc U$ is a pure rotation
\[ \vc U=r\sin\theta\,\Omega\,\vc e_\phi   \]
and introduce the shear vector
\[ \hat{\vec\Omega}=r\sin\theta\,\nabla\Omega   \]
In the limit $r\sin\theta/H_B\rightarrow\infty$ the form of the vector operators
simplifies to their form for the local Cartesian frame $\vc e_{x'}=\vc
e_\theta$, $\vc e_{y'}=\vc e_\phi$, $\vc e_{z'}=\vc e_r$. Now we introduce a
new frame $xyz$ by rotating $x'y'z'$ around $\vc e_\phi$ with an angle 
$-\pi/2\le\Delta\theta <\pi/2$ so that $\hat\Omega_x=0$. With these assumptions
and notations the poloidal and toroidal parts of equation (\ref{eq:avind}) read
\< \ptl_t A=\alpha B+\beta\,\nabla^2 A   \label{eq:pol} \>
\< \ptl_t B={\hat\Omega}_z \,\ptl_x A -\alpha\nabla^2 A +\beta\,\nabla^2 B 
   \label{eq:tor}  ,  \>
known as the classic dynamo equations. 

It is clear from (\ref{eq:pol}--\ref{eq:tor}) that the role of the
pseudoscalar $\alpha$ is to turn the poloidal and toroidal field components
into each other which implies some kind of helical motion. The classic 
candidate for this, suggested by \citeN{Parker:cyclonic} is the passive
advection of fields by helical convective motions. Later, alternative
mechanisms for an $\alpha$-effect were also proposed, based on a dynamic
interaction of field and motions (see Section 4.3 below). A general property of
these mechanisms is that $\alpha$ turns out to be positive in the bulk of the
solar convective zone in the northern hemisphere, while it tends to be negative
in the stably stratified layer below. (Being a pseudoscalar, $\alpha$ changes
sign between hemispheres.) 

The shear $\hat\Omega$, associated with differential rotation, in turn, winds
up the poloidal field into a toroidal component. Without the $\alpha$ and
$\Omega$ terms we would be left with a diffusively decaying field, so at least
one of these terms is necessary for dynamo action in both equations. Depending
on which, if any, of the dynamo terms in (\ref{eq:tor}) is discarded, we
distinguish $\alpha^2$, $\alpha\Omega$ and  $\alpha^2\Omega$ dynamos.
$\alpha^2$ dynamos can be shown to give rise to non-oscillatory behaviour and
toroidal and poloidal field amplitudes of the same order of magnitude which
does not agree with the properties of the solar dynamo. Thus, in what follows
we will concentrate on $\alpha\Omega$ dynamos, neglecting the second term on
the r.h.s. of (\ref{eq:tor}). (Note, however, that under more general
conditions than those considered here, oscillatory $\alpha^2$ dynamos can also
be constructed, as pointed out recently by \citeNP{Schubert+Zhang}.)

Assuming that beside $\alpha$ and $\beta$, ${\hat\Omega}_z$ can also be
regarded constant, the system (\ref{eq:pol}--\ref{eq:tor}) is homogenous and
linear, admitting wavelike solutions of the form
\< B=B_0 \exp\,[i(\omega t -\vc k\vc x)]   \label{eq:Bansatz} \>
\< A=A_0 \exp\,[i(\omega t -\vc k\vc x+\delta)]  \label{eq:Aansatz}  \>
where $\omega=\omega_R+i\omega_I$ is a complex frequency while all other
variables are real. $\omega_R$, $A_0$ and $B_0$ can be taken to be non-negative
without loss of generality. Introducing the (signed) Reynolds numbers
\< R_\alpha=\alpha/\beta k \qquad R_\Omega={\hat\Omega}_z/\beta k^2 \>
as well as the dynamo number and the nondimensional frequency
\< D=R_\alpha R_\Omega \qquad \tilde\omega=\omega/\beta k^2  \label{eq:Ddef} \>
and substituting the {\it Ansatz} (\ref{eq:Bansatz}--\ref{eq:Aansatz}) into
(\ref{eq:pol}--\ref{eq:tor}) we find
\< (1+i\tilde\omega)A_0=k^{-1}R_\alpha B_0\, e^{-i\delta}  \label {eq:Awave} \>
\< (1+i\tilde\omega)B_0=-ik_x R_\Omega A_0\, e^{i\delta}  
    \label {eq:Bwave}  \>
The product of these latter equations is
\< (1+i\tilde\omega)^2=-iDk_x/k    \label{eq:waveprod}  \>

As only $k_x$ appears in (\ref{eq:Bwave}), no  unstable modes (self-excited
field  or ``dynamo waves'') exist with $k_x=0$. Remembering the way we oriented
our  $x$-axis this implies that {\it dynamo waves propagate along isorotational
surfaces.} But in which direction?  The imaginary part of (\ref{eq:waveprod})
reads
\< 2\tilde\omega_R (1-\tilde\omega_I)=-Dk_x/k   \label{eq:waveprodim} \>
As for unstable modes (self-excited field or ``dynamo waves'') $\omega_I\le 0$,
(\ref{eq:waveprodim}) yields the important result
\< k_x D <0 , \>
known as the {\bf Parker--Yoshimura sign rule.} Thus, e.g.\ in the northern
hemisphere equatorward propagation ($k_x>0$) implies $D<0$. With a positive
$\alpha$, as is the case in the bulk of the convective zone, this implies
$\ptl_r\Omega<0$, an outward decreasing rotational rate. This was indeed the
general expectation for the solar internal rotational law in the 1960's and
1970's.

The solution of (\ref{eq:waveprod}) is
\< \tilde\omega=i\pm (1-i)(-Dk_x/2k)^{1/2}  \label{eq:waveprodsol}  \>
For unstable modes obviously the plus sign applies. So the growth rate is
\< -\tilde\omega_I=(-Dk_x/2k)^{1/2}-1    \label{eq:growthrate}  \>
Unstable modes thus exist when $|D|\ge 2$ in nondimensional units. As according
to equation (\ref{eq:Ddef}) $|D|$ decreases with $k$, it is the {\bf lowest $k$
modes,} with a scale comparable to the solar radius $R_\odot$ that have the
highest growth rate and {\bf will dominate\/} the solution. (Note that this
implies that our formalism, derived for the limit
$kr\sin\theta\rightarrow\infty$, is strictly speaking invalid for these modes
---nevertheless it may still be used for general guidance.)

Note that when $\tilde\omega_I=0$, $\tilde\omega_R=1$ follows from
(\ref{eq:waveprodsol}): {\bf the period of the critical mode is thus just the
diffusive timescale corresponding to $R_\odot$.} Estimating $\alpha$ and 
$\hat\Omega_z$ on the basis of helical convection and the observed differential
rotation, $D$ proves to be order of unity for a convection zone dynamo, showing
that the dynamo is indeed approximately critical, and thus naturally explaining
the agreement of the cycle period with the diffusive timescale.

Finally, let us note that it is straightforward to work out from the above
formulae that for an equatorward propagating wave, the phase of the radial
field component relative to the toroidal field is $-3\pi/4$ if $\alpha>0$ in
the northern hemisphere and $\pi/4$ otherwise (\citeNP{Stix:phase.dilemma}). 

Taken altogether, the above considerations showed that for the expected
positive $\alpha$-effect in the solar convective zone, assuming an inwards
increasing rotational rate, one can correctly reproduce the cycle period, the
equatorward branches of the butterfly diagram as a dynamo wave, and the
low-latitude phase relationship. (The high-latitude poleward branch could
obviously be reproduced by assuming $\ptl_r\Omega>0$ there, though this line
was not pursued, relying on the Babcock-Leighton approach instead.) All this
gave the impression that, missing details apart, the basic mechanism of the
solar dynamo is well understood.

\alszekcio{2.2. Crisis}
The first warning signs that something is amiss began to appear towards 1970,
with the realization that most of the magnetic flux in the solar photosphere,
and presumably below, is present in a strongly intermittent form, concentrated
into strong flux tubes (\citeNP{Weiss:first}, \citeNP{Sheeley:strongfield}, 
\citeNP{Stenflo:firstkG}, \citeNP{Howard+Stenflo}). Flux tube theory developed 
in the 1970's and it became clear that for thicker flux bundles, owing to their
lower surface/volume ratio, volume forces such as buoyancy, curvature and
Coriolis forces dominate over the drag of the surrounding plasma flows acting
on the surface. These tubes can then move largely independently of the
surrounding flows, invalidating simple one-fluid descriptions like mean field
theory. Thus, solar magnetic fields can be divided into two components: passive
fields, consisting of thin flux fibrils, that move passively with the flow
owing to the drag and are the subject of mean field theory; and active fields,
consisting of thick flux bundles moving under the action of volume forces. And
the characteristics of large active regions strongly suggested that they are
essentially (fragmented) loops formed on toroidal flux bundles of $10^{22}\,$Mx
which clearly fall in the active category, outside the jurisdiction of mean
field models. 

What is more, \citeN{Parker:buoy.prob} called attention to the fact that such
flux bundles cannot be stored in the convective zone for a time scale
comparable to the cycle period, being subject to buoyant instabilities that can
rapidly remove the whole tube from the zone. The only place to store these
tubes is near the bottom of the convective zone, especially in the stably
stratified but still turbulent {\it lower overshoot layer} below it. Toroidal
flux tubes lying here may still develop finite-wavelength buoyant instabilities
that may give rise to loops erupting through the convective zone into the
atmosphere, producing active regions. This scenario has gained firm foundations
with the first nonlinear calculation of the emergence of such loops through the
convective zone (\citeNP{FMI:classic}), and such {\it flux emergence models}
have by now evolved into an independent chapter of the global dynamo problem.
In the present review we will not deal with this topic in detail (see the
review by \citeNP{FMI:Freibg}), even though flux emergence models are the only
real ``success story'' of dynamo theory since the 1960's. While many details
are still unclear, by now these models can reproduce sunspot proper motions and
active region flux distributions to a quite convincing detail. A very robust
main conclusion from the models, of great importance for the global dynamo,
is that in order to reproduce the observed characteristics of active regions
the toroidal flux tubes must have a field stregth of about $10^5\,$G ---an
order of magnitude higher than the turbulent equipartition field in the deep
convective zone. Explaining the origin of such strong fields is a major
challenge for dynamo theory.

Guided by these realizations, in the 1980's the first attempts were made to
construct dynamos operating in the lower overshoot layer. The unknown profiles
of $\alpha$ and the differential rotation, however, greatly impeded progress,
allowing a far too wide parameter space to play with. Therefore, attempts were
made to numerically simulate the whole convective zone, with a consistent
picture of differential rotation, helical convection, and dynamo
(\citeNP{Glatzmaier:badsimu}). Nevertheless, the results (poleward propagating
dynamo waves) were at odds with the  observations, and when finally even the
predicted differential rotation profile (constant on cylinders) was proven
wrong by helioseismic measurements (constant on cones), the simulational
approach was abandoned (cf.\ the remarks in Section~5).

On the other hand, the helioseismic determination of internal differential
rotation gave new impetus to mean field dynamo theory. Those inversions clearly
showed that most of the shear (the $\Omega$-effect) is concentrated in a thin
layer near the bottom of the convective zone, known as the tachocline. This was
seen as further evidence that a thin layer situated about $200\,000\,$km below
the solar surface is of key importance for the working of the solar dynamo.
Depending on the physical viewpoint we study it from, this layer is
alternatively called dynamo layer, overshoot layer or tachocline. In the
present review for simplicity I will refer to it as the DOT (Dynamo ---
Overshoot --- Tachocline) layer.

\begin{table}[htb]
\caption{\small\it
A classification scheme for solar dynamo models
\label{table:classif} }
\medskip
\centerline{\psfig{figure=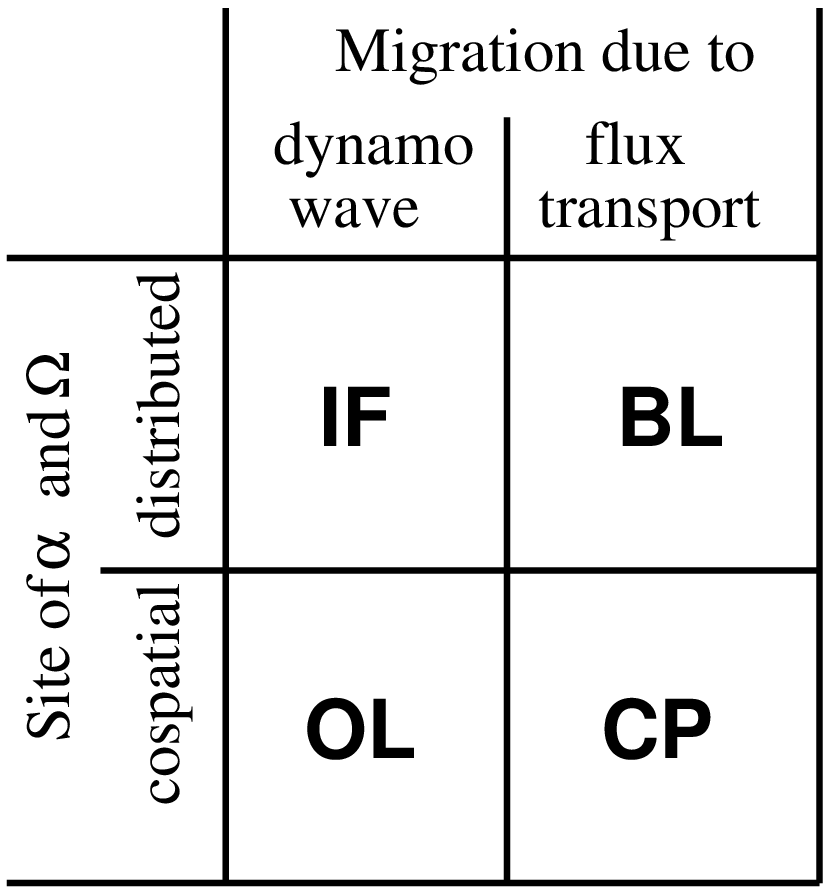,width=5 cm}}
\vskip -0.3 cm
\end{table}

\begin{figure*}[tbp]
\centerline{{\psfig{figure=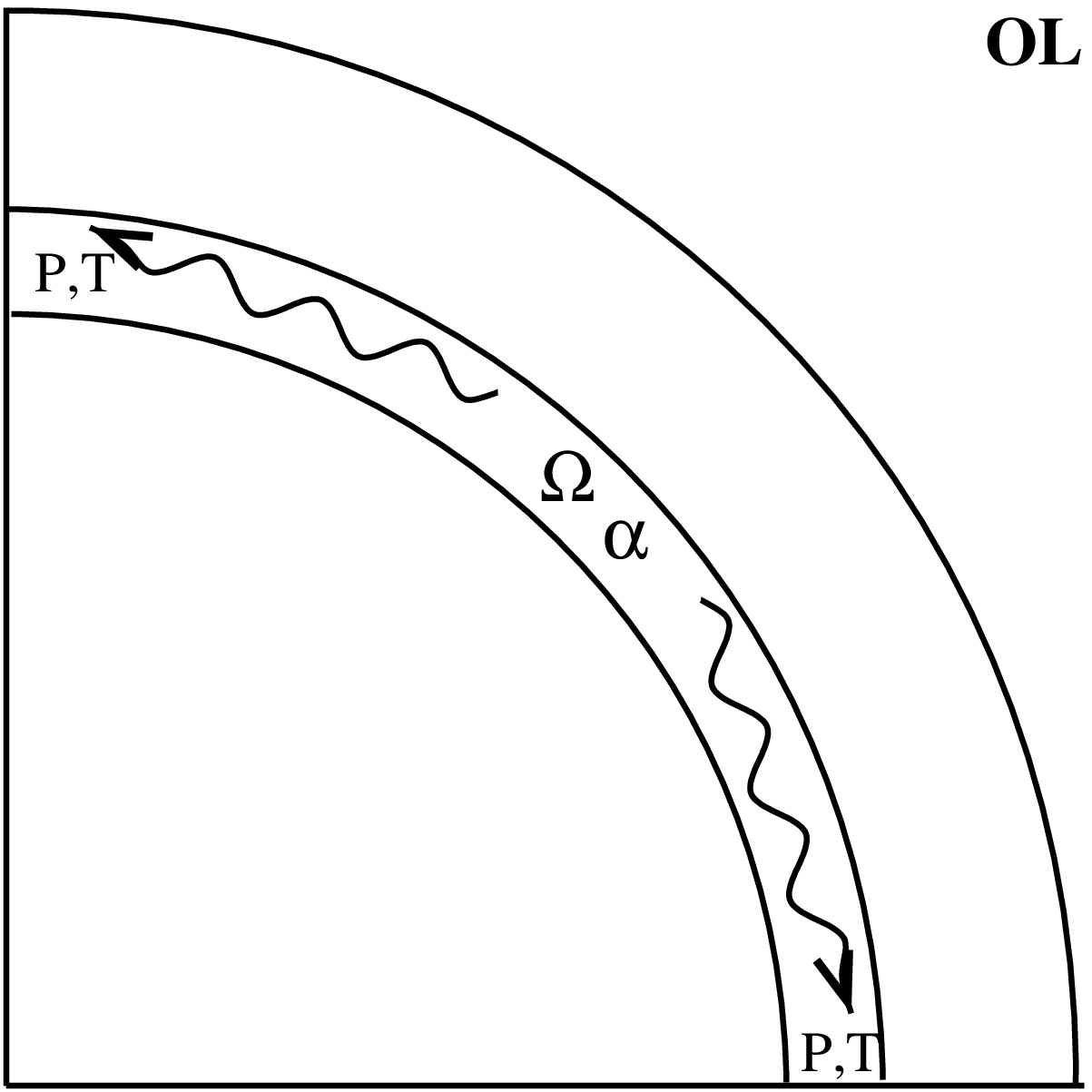,width=7.5 cm}}{\hfill}{\psfig{figure=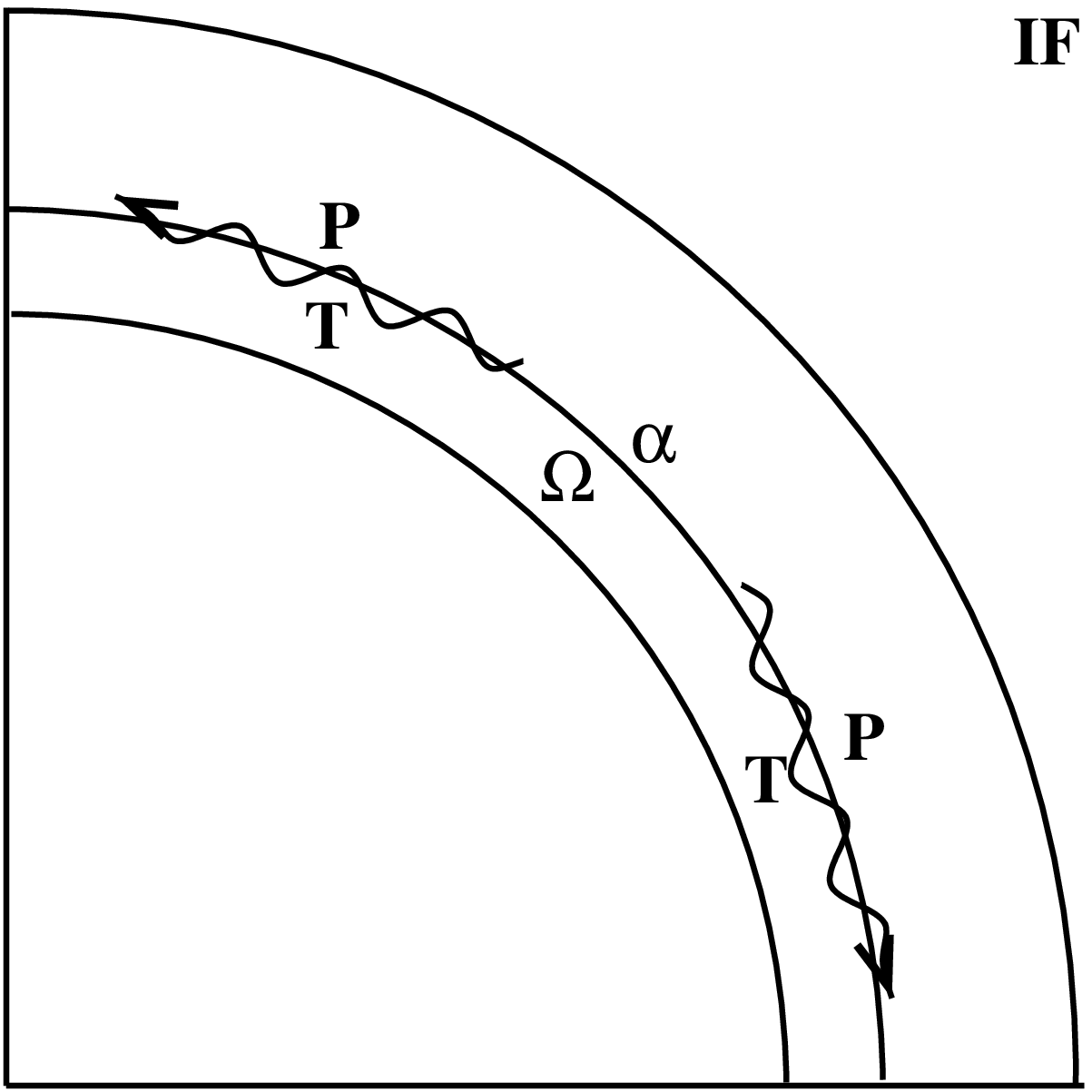,width=7.5 cm}}}
\bigskip
\centerline{\psfig{figure=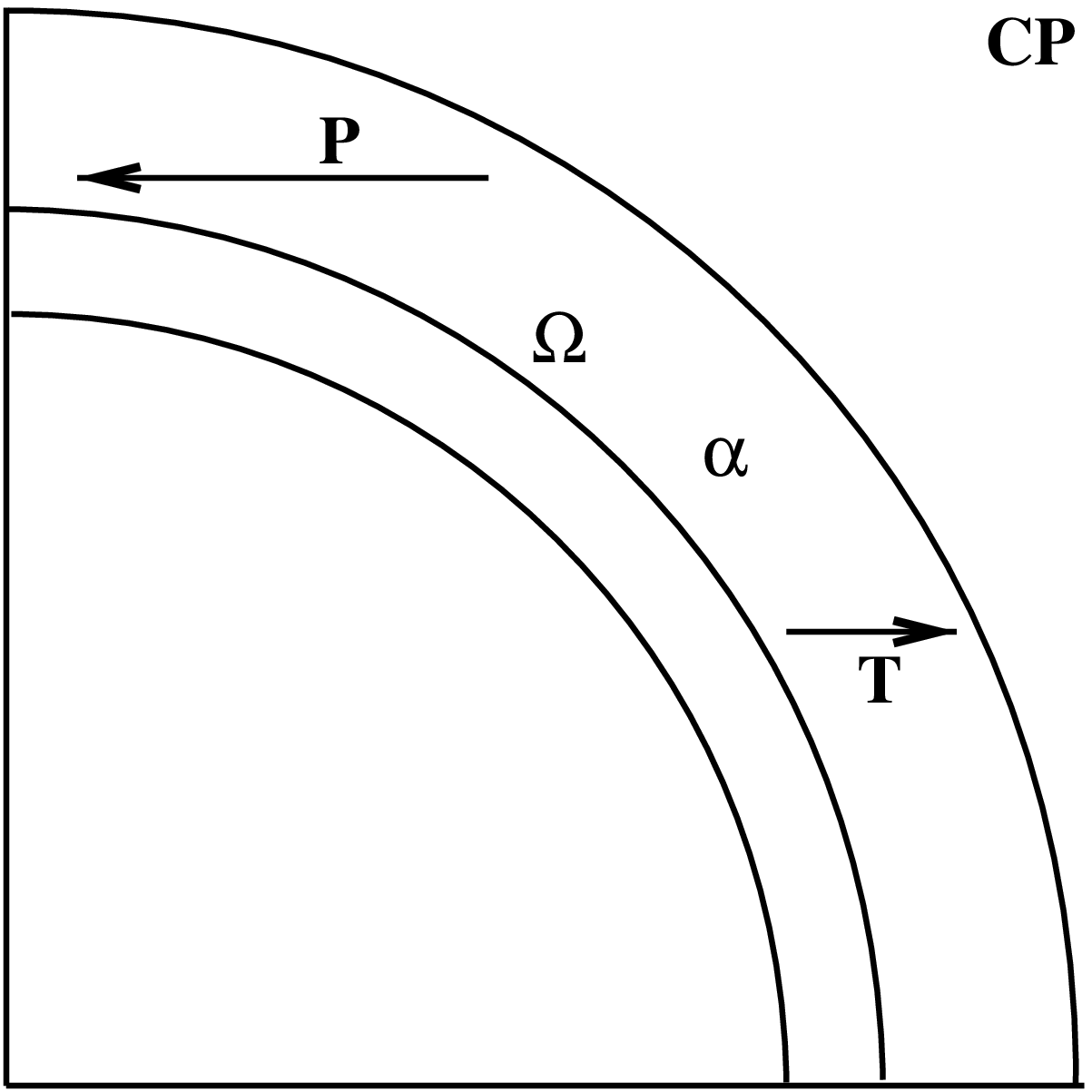,width=7.5 cm}{\hfill}\psfig{figure=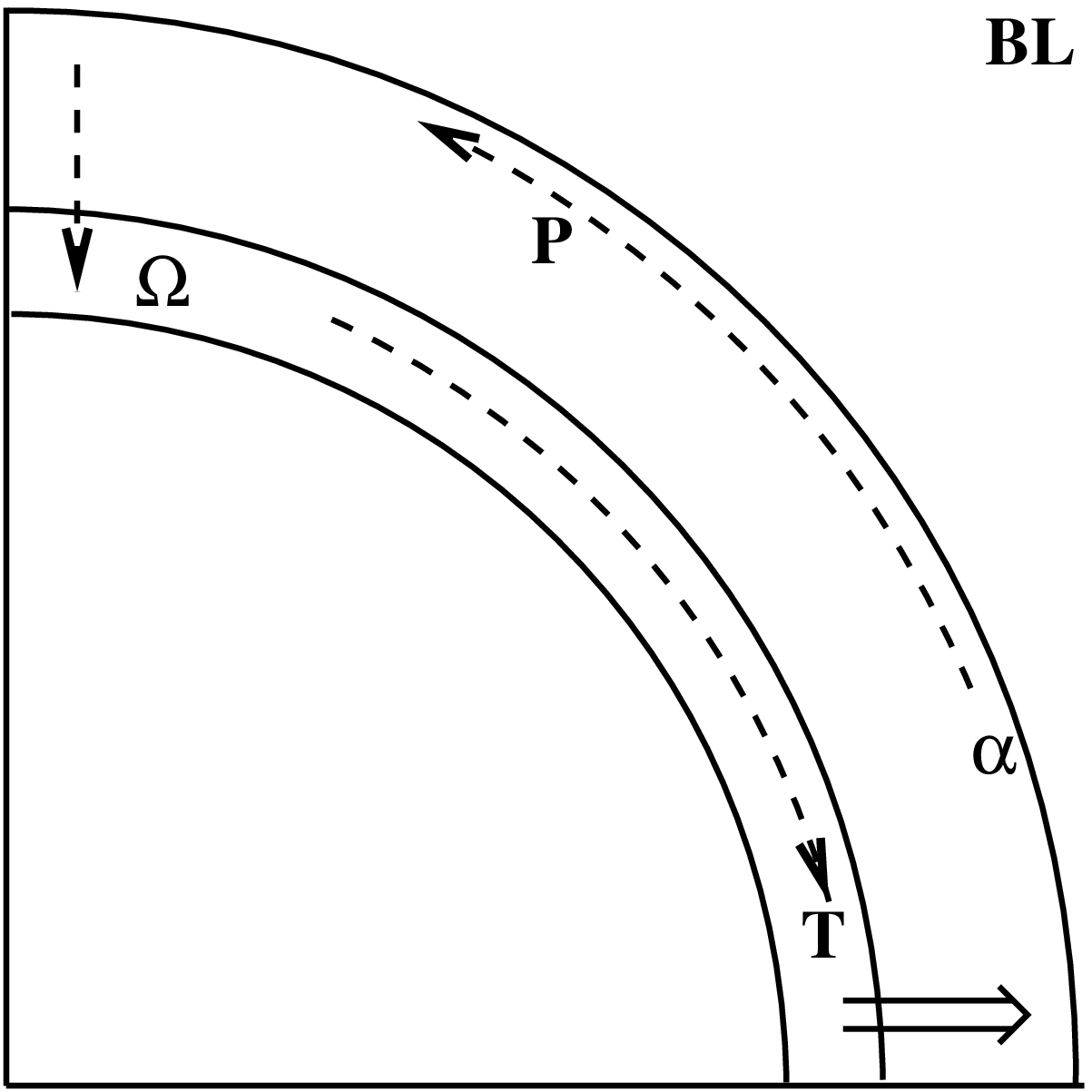,width=7.5 cm}}
\bigskip
\centerline{\psfig{figure=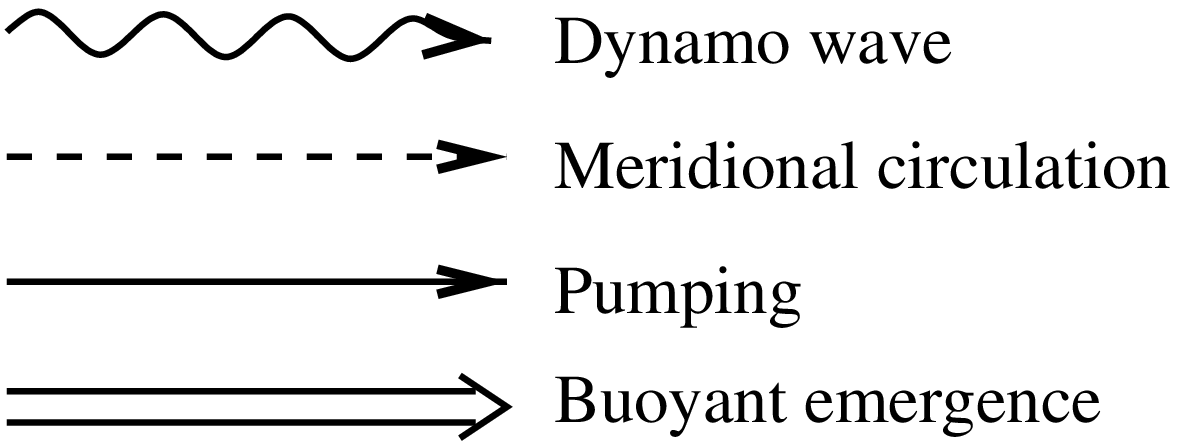,width=7.5 cm}}
\caption{\small \it
Schematic illustration of the main features of main solar dynamo families. The
sites of $\alpha$ and $\Omega$ effects and toroidal/poloidal field transport
mechanisms and directions are shown in a meridional quadrate. The internal
layers marked schematically are, from the inside outwards, the radiative, DOT
and convective layers.
{\bf OL}: Overshoot layer dynamos;
{\bf IF}: Interface dynamos;
{\bf CP}: Cospatial pumping models;
{\bf BL}: Babcock-Leighton type models;
\label{fig:classif} }
\end{figure*}

\szekcio{3. MAIN FAMILIES OF MODELS}
Solar mean field dynamo theory in the past decades gave rise to a bewildering
variety of models. Nevertheless, all recent models (i.e.\ those using the
helioseismic rotation law) can be classified into just four main types
according to a plausible classification scheme (Table~\ref{table:classif} and
Figure~\ref{fig:classif}). One classification parameter here divides the models
according to whether they still interpret the butterfly diagram as a dynamo
wave just like the orthodox convection zone dynamos did, or if they
substitute that with some flux transport mechanism (meridional circulation or
pumping). The other parameter in turn divides the models according to whether
the $\alpha$- and $\Omega$-effects are cospatial or ``distributed'', i.e. they
take place in different (adjacent or very distant) parts of the model volume.
The resulting four model types are as follows.

\bigskip

\alszekcio{3.1. Cospatial wave models: OL dynamos\vspace{-0.5 cm}}
Widely known as ``overshoot layer'' or OL dynamos, these are perhaps the most 
conservative models that simply replant the concepts of the convective zone
dynamos of the 1960's and 70' into the DOT layer. As the helioseismic
inversions show $\ptl_r\Omega>0$ at low latitudes, these models need to assume
$\alpha<0$ to get the right migration directions. This assumption is rather
plausible in the DOT layer for several different physical mechanisms for
$\alpha$. The state-of-the-art in this approach is represented by the model of
\citeN{Rudiger+Brandenburg}.

{\it Successes: }

\begin{lista}

\item The butterfly diagram comes out right. The Parker-Yoshimura rule
leads to polar and equatorial branches separated at the corotational latitude,
as observed.

\end{lista}

{\it Difficulties:}

\begin{lista} 

\item The low-latitude phase dilemma: as now $\alpha<0$, the radial field is
found to be nearly in phase with the toroidal field, instead of being in
antiphase as observed. 

\item The cycle period tends to be too short for thin layer models. This is
basically because the same amount of differential rotation is now concentrated
in a much thinner layer, leading to much stronger shear and much higher dynamo
numbers.  This may be compensated by reducing $\alpha$, e.g.\ by arguing that,
as nonlinear effects act via a quenching of the $\alpha$ and $\Omega$
mechanisms, it is natural to expect that after saturation the dynamo will be
effectively critical. Yet the degree to which the nonlinearity can increase the
cycle period depends a lot on the assumptions made, and in general it does not
seem sufficient (\citeNP{Rudiger:Freibg}). At any rate, certain ``tricks'',
such as using an $\alpha^2\Omega$ dynamo (\citeNP{Gilman+:layer.dyn}) or 
introducing an intermittence factor (\citeNP{Rudiger+Brandenburg}), can save
the models, but then the crude coincidence of the period with the lateral
diffusive timescale is coincidental.

\item For an equatorial confinement of strong fields the $\alpha$-effect needs
to be arbitrarily confined to low latitudes. Nevertheless, such
$\alpha$-distributions are indeed found in some calculations (cf.
\citeNP{Schmitt:Potsdam}).

\end{lista}

\alszekcio{3.2. Distributed wave models: IF dynamos}
\citeN{Parker:interface} suggested a dynamo where the diffusivity
discontinuously varies by many orders of magnitude across a surface. The
$\alpha$-effect operates on the high-diffusivity side of the interface, while
the $\Omega$-effect (shear) is limited to the low-diffusivity side. He showed
that under these conditions a dynamo wave can be excited, obeying the
Parker-Yoshimura sign rule. The attractive feature of this model is that the
toroidal field generated on the low diffusivity side can be made arbitrarily
strong by reducing the value of magnetic diffusivity there. Thus, the origin of
$10^5\,$G fields could be explained.

Parker's analytic, plane parallel model has been extended to more realistic
situations and incorporated in full solar dynamo models in a number of
papers (\citeNP{Charbonneau+McGregor:IFdynamo}, \citeNP{Markiel+Thomas}).
Unfortunately, the results are somewhat contradictory owing to numerical
problems related to modelling the discontinuity. In this context we may perhaps
note that the main physical difference of these IF models compared to the OL
models is the spatial separation of $\alpha$ and $\Omega$. The introduction of
a discontinuity between them is an added feature that can simplify the analytic
treatment but at the same time complicate the numerical calculations. A model
where the diffusivity is continuously distributed, albeit with a sharp
gradient, with $\alpha$ and $\Omega$ concentrated on the two sides of this
gradient, may well be worth considering. Beside being more realistic, such a
model might also avoid the numerical problems mentioned.

{\it Successes:}

\begin{lista}
\item Strong toroidal fields can be readily explained.

\end{lista}

{\it Difficulties:}

\begin{lista}
\item At the present stage, numerical difficulties prevail. An evaluation of 
the physical performance of the models will only be possible after these are
resolved.

\end{lista}

\alszekcio{3.3. Cospatial transport models: CP dynamos}
In an inhomogeneous medium, the proper expression of $\vec{\cal E}$ is more
general than equation (\ref{eq:Ekif}), the scalars $\alpha$ and $\beta$ being
substituted by tensorial expressions. A tensorial $\mx\alpha$-term in 
(\ref{eq:Ekif}) can then alternatively be written as
\[ \mx\alpha\vc B=\alpha\vc B+\mx\alpha_S\vc B -\vec\gamma\times\vc B  \]
where $\alpha=\mbox{Tr\,}\mx\alpha$, the last term is the vectorial product
equivalent of the action of the  antisymmetric part of $\mx\alpha$ on $\vc B$,
and $\mx\alpha_S$ is the symmetric and traceless part of $\mx\alpha$.
Substituting (\ref{eq:Ekif}) into (\ref{eq:avind}) we see that the role of
$-\vec\gamma$ is analoguous to that of $\vc U$, i.e.\ it formally describes an
advection of the magnetic field. This effect is called the (normal) {\it
pumping\/} of the field along the inhomogeneity. The $\mx\alpha_S$-term can be
shown to give rise to a similar effect with the difference that the sign of the
transport depends on the orientation of the field component perpendicular to
the pumping direction (anomalous pumping; see \citeNP{Petrovay:NATO} for a
detailed discussion). 

It must be stressed that the descriptions using a tensorial alpha and those
using pumping effects are formally equivalent. The advantage of the pumping
formalism is that it helps one to get a physical ``feeling'' of the processes
at work.

Depending on the particular inhomogeneity associated with the pumping, one can
speak of density pumping, turbulent pumping etc. It was suggested by
\citeN{Krivod+Kichat} that, instead of appealing to a dynamo wave, the field
migration patterns can also be interpreted by density pumping, directed towards
the rotational axis for poloidal fields and away from it for toroidal fields.
This pumping is supposed to operate throughout the convective zone, instead of
being confined to the DOT layer. A more detailed model along these lines has
recently been constructed by \citeN{Kichat+:pump.dynamo}. While the model
reproduces well the phase relation for the torsional oscillations, it does not
even address questions such as the origin of the deep-seated toroidal field.
Thus, while the basic concept is interesting, at the present stage this
``cospatial pumping'' approach cannot be regarded as a serious aspirant for the
explanation of the global dynamo.

\bigskip 
\alszekcio{3.4. Distributed transport models: BL dynamos\vspace{-0.5 cm}}
These are generally known as ``Babcock--Leighton-type'' models (hence the BL
code). While they indeed grew out of the semiempirical approach of
\citeN{Babcock:merid.circ} and \citeN{Leighton:diffusion} to the solar cycle,
they did get a lot more radical in the 1990's. Indeed, in the era of convective
zone dynamos, in the 1960's and 70's, the Babcock-Leighton approach was not
seen as necessarily conflicting with the dynamo wave theory of the cycle. With
a dynamo operating throughout the convective zone, the $\alpha$-effect caused
by the inclination of active region axes relative to E--W (which is the
cornerstone of this approach) could be considered as helical convection caught
in the act; and in his mathematical formulation of the model
\citeN{Leighton:diffusion} explicitly assumed that the toroidal field migration
is the result of a dynamo wave. It was only the poleward migration of the
poloidal field that was interpreted in terms of a lateral transport.

In subsequent versions of the model, meridional circulation plays the main role in
transporting the poloidal fields to the poles near the surface. And from here
it was just one step to close the circle and assume that the deep return flow
of meridional circulation is responsible for the equatorward drift of the
toroidal field. This step, made by \citeN{Wang+:1.5D} was radical indeed: it
represented the first break with the canonical interpretation of the butterfly
diagram as a dynamo wave, generally accepted for three decades. 

The model essentially works like a conveyor belt: the poleward meridional
circulation near the surface transports the poloidal fields towards the poles
at high latitudes, giving rise to the poleward branch of the butterfly diagram.
At the poles, the fields are advected down to the bottom of the convective zone
where the shear converts them into toroidal fields that get amplified while
advected towards the equator. Once these are strong enough, they are supposed
to form buoyantly emerging loops at low latitudes that give rise to active
regions, the Coriolis force lending an inclination to the loop planes, i.e.\
introducing an $\alpha$-effect, and thus regenerating the poloidal field.

More recent versions of the model (\citeNP{Choudhuri+:mixed.transp},
\citeNP{Dikpati+Charbonneau}) have developed it into an internally consistent
modelling approach with the ambition of yielding a complete description of the
solar cycle. As such, this family of models is at present the most serious
competitor of the OL class.

{\it Successes:}

\begin{lista}

\item The low-latitude confinement of strong activity comes out rather
naturally. (The toroidal field is amplified by shear as it is advected
equatorward.) 

\item A tolerable reproduction of the extended butterfly diagram (although the
latitude where the two branches part tends to be too high). Note, however, that
the original (one-dimensional) Babcock--Leighton models gave a much closer fit
to the observed migration patterns, which was their main asset. In the
extension to two dimensions, BL dynamos had to sacrifice this achievement.

\end{lista}

{\it Difficulties: }

\begin{lista}

\item The agreement of the corotational latitude with the latitude where the
butterfly branches diverge must be considered coincidental (if reproduced at
all).

\item The $\alpha$-effect is confined to the surface. This is probably
unrealistic even for flux tube alpha, cf.\ the discussion in Section~4.3 below.

\item The model, as all models relying on a flux tube alpha, is not
self-exciting, as this $\alpha$-effect works for strong fields only. It needs 
another dynamo mechanism to ``kick it in''.

\item A serious problem is that the approach only works with an unrealistically
low value for the turbulent diffusivity in the convective zone, $\beta\sim
10\,$km$^2/$s. (This is needed to keep the two parts of the ``conveyor belt''
separated. In fact the main difference between the BL and IF models is that the
separated seats of $\alpha$ and $\Omega$ communicate by diffusion in the IF
models.) There is no justification for such a low value. On physical grounds,
$\beta\sim 0.5 lv\simeq 500\,$km$^2/$s is expected ---indeed, even the
one-dimensional Babcock--Leighton models predicted $600\,$km$^2/$s. According
to the controversial proposal of \citeN{Vainshtein+Rosner:first}, on the other
hand, the diffusivity should have been quenched by {\it many} orders of
magnitude to values far lower than those used in the BL models. (Note that by
now this question has been settled: diffusivity suffers no strong quenching in
realistic, intermittent fields in three dimensions,
\citeNP{Gruzinov+Diamond:letter}, \citeNP{Petrovay+Zsargo}.) 

\begin{figure}[htb]
\centerline{\psfig{figure=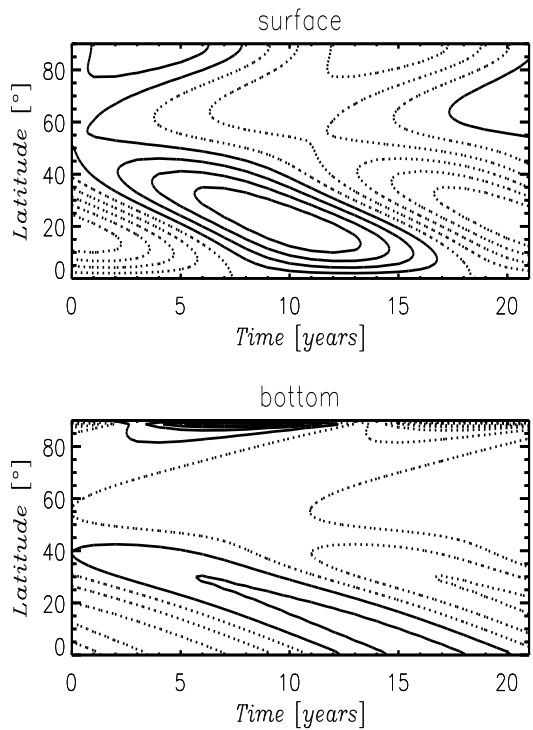,width=\columnwidth,height=10 cm}}
\vskip -0.5 cm
\caption{\small \it
Steamy window: with realistic turbulent transport parameters, an arbitrary 
time-dependent poloidal field pattern imposed at the bottom of the convective
zone is reflected, somewhat blurred, at the surface. After 
Petrovay \& Szak{\'a}ly (1999)
\label{fig:steamy} }
\end{figure}

One may wonder if the effect of the strong diffusive link can be reduced by
some kind of selective field pumping that would keep the two field components
apart. \citeN{Petrovay+Szakaly:2d.pol} investigated this possibility in a model
of poloidal field transport, incorporating all known transport effects
(circulation, diffusion, turbulent pumping, density pumping). It turns out that
diffusion is dominant, forging such a strong link between the DOT region and
the surface that any migration pattern imposed at the bottom will be reflected
at the top (Fig.~\ref{fig:steamy}). This seems to represent a serious 
difficulty for BL models.

\end{lista}

\szekcio{4. KEY ISSUES}
\alszekcio{4.1. Structure of the DOT region\vspace{-0.8 cm}}
It is obvious from the above model descriptions that a thorough understanding
of the structure and dynamics of the DOT layer is crucial for advance in solar
dynamo theory. Unfortunately, at present we do not have a reliable model for
this layer. As a complicated interplay of convection, rotation and magnetic
fields is expected there, as a first step one would expect to develop models
for just one aspect of the problem: a model for overshoot with no rotation and
magnetic fields; a non-magnetic model for the tachocline with no overshoot;
etc. Nevertheless, we lack even such simplistic models, and surprisingly little
effort is made to construct them. 

In the case of overshoot, the way to construct such a model is in principle
well known: it consists in solving a conveniently truncated hierarchy of the
Reynolds stress equations. \citeN{Marik+Petrovay:SOGO} present preliminary
results from a numerical study of this problem.

Helioseismic inversions show that the thickness of the tachocline is about
$0.1\,R_\odot$. This is puzzling, as even in the absence of turbulent
diffusion, the Eddington--Sweet circulation should have mixed these layers
enough in $4.6\cdot 10^9$ years to extend the tachocline to much greater depths
(\citeNP{Spiegel+Zahn}). It is most often assumed that the solution to this
``thin tachocline problem'' resides in anisotropic turbulence: an extremely
strong horizontal anisotropy would indeed effectively smooth out horizontal
gradients without contributing much to the vertical transport. But in view of
strong nonlinear mode coupling in turbulence, such an extreme anisotropy seems
dubious. \citeN{Dajka+Petrovay:SOGO} explore the alternative possibility that
the necessary horizontal momentum transfer is due to an inverse
$\Lambda$-effect: however, the necessary amplitude of $\Lambda$ proves even
more unrealistically large. Magnetic decoupling of the envelope from the
radiative interior is a third explanation put forward by
\citeN{Rudiger+Kichat:thin.tacho}.

\alszekcio{4.2. High-latitude field patterns}
In order to be able to decide between various dynamo models, one needs an issue
on which they yield conflicting predictions that are relatively easy to test.
One such case is the migration of the toroidal field at high latitudes. In
dynamo wave (OL and IF) models at high latitudes all field components migrate
polewards, following the Parker--Yoshimura rule. In contrast, the toroidal
field in BL  models typically shows an equatorward migration, advected by the
return flow of meridional circulation near the bottom of the convective zone.
Thus, if a clear signature of  migrating high-latitude toroidal fields were
found, this could solve the  problem in either way, depending on the direction
of migration. It is often claimed (e.g.\ \citeNP{Harvey:NATO}) that high 
latitude ephemeral active regions show an  equatorward migration. But a closer
examination of such claims shows that they are based only on the
fact that high latitude ephemeral regions as a whole tend to lie in the
backwards extension of the low-latitude butterfly ``wings'' (just as polar
faculae do), and not on a detailed study of migration patterns {\it among\/}
high-latitude regions only.  On the other hand, \citeN{Callebaut+Makarov} claim
that at least 50\,\% of polar faculae (well known for their poleward drift)
correspond to dipoles with a preferential east--west orientation, thus forming
part of the toroidal  field. It has even been claimed  that the
highest-latitude part of the sunspot butterfly diagram also shows a  poleward
drift (\citeNP{Becker:poleward}). A clarification of this issue would clearly
be important. 

A related problem is the origin of ephemeral active regions in general. If they
are used as magnetic tracers we would obviously like to know what kind of field
do they trace?

\bigskip

\alszekcio{4.3. Origin and profile of $\alpha$\vspace{-0.7 cm}}
It is sometimes claimed that the origin of the $\alpha$-effect assumed is a
basic difference between various dynamo models, so that excluding a certain
$\alpha$-mechanism amounts to excluding a dynamo variety. In reality, our
knowledge of all possible types of $\alpha$-effects is so scarce that wildly
differing $\alpha$-profiles can be derived for the same mechanism, depending on
particular modelling assumptions. And vice versa: a given profile, used as
input in a dynamo model, may be the consequence of various $\alpha$-mechanisms.

More in-depth study of the possible $\alpha$-mechanisms that can reduce this
uncertainty could nevertheless really be used to constrain dynamo models in the
future. For now, I will argue that, independently of its mechanism, $\alpha$
should be expected to be concentrated towards the bottom of the convective
zone, or at least to pervade it uniformly (and not to be concentrated to the
surface, as assumed in BL models). Let us see this for the 
four $\alpha$-mechanisms that have been proposed.

\begin{lista}

\item {\bf Cyclonic convection.} In this classic variety of the
$\alpha$-effect, proposed by \citeN{Parker:cyclonic}, $\alpha$ results from
the effect of rotation on convection, so it increases with the Coriolis
number, peaking near the bottom of the solar convective zone. Its sign is
positive in the unstable layer and negative in the overshoot layer below.

\item {\bf Magnetostrophic waves.} This mechanism was proposed by 
\citeN{Schmitt:mstroph} specifically to suggest an alternative
$\alpha$-mechanism for the overshoot layer that will not be suppressed by the
strong toroidal fields there. It consists in growing helical waves and its sign
is negative, as required by the dynamo wave models. More detailed calculations
of its profile were performed by \citeN{AFM+Schussler:mstroph}, though the
results may rely too heavily on the background stratification of the
(incorrect) overshoot model used.

\item {\bf Flux loop alpha.} The magnetix flux loops emerging through the
convective zone and creating the active regions are essentially large-amplitude
nonlinear versions of the above mentioned unstable magnetostrophic waves,
giving rise to a similar (but positive) $\alpha$-effect. It is clear that after
their decay, owing to their tilt, active regions contribute to the passive
poloidal field. But is this contribution limited to the surface, or do the
``feet'' of the loop similarly decay, down to the bottom of the convective
zone? This problem relates to the next subsection, but one expects that the
eroding action of external turbulence (\citeNP{Petrovay+FMI:erosion}) will lead
to the decay of the whole loop, thereby extending the $\alpha$-effect down to
its footpoints. In fact, our turbulent erosion calculations show that 90\% of
the magnetic flux in an emerging flux loop is lost {\it before\/} it reaches
the surface! This lost flux, already submitted to the action of Coriolis force,
should contribute to the poloidal field and to the $\alpha$-effect far more
than the actual active regions. And then we did not even mention the
possibility of ``failed active regions'': flux loops that, their field
strengths being too weak, never make it to the surface
(\citeNP{Petrovay+Szakaly:AA1}, \citeNP{FMI+:explosion}). All this strongly
speaks for a flux loop $\alpha$ concentrated to the bottom of the convective
zone.

\item{\bf Unstable Rossby waves.}
Recently, \citeN{Dikpati+Gilman} considered a shallow-water model of the 
tachocline, analysing the behaviour of small perturbations of its  geostrophic
equilibrium state. They find that such large-scale perturbations are unstable
if the subadiabaticity is low enough and they are characterized by a
correlation between the vertical components of velocity and vorticity, i.e.\ by
a non-vanishing mean helicity. In other words, these perturbations essentially
behave like global-scale unstable Rossby waves, and owing to their mean
helicity they can be expected to yield an $\alpha$-effect.

\end{lista}

\alszekcio{4.4. The ultimate fate of emerged flux}
This problem is interesting in its own right, as well as because of its
importance for the $\alpha$-effect problem. Once the active region has decayed,
what happens to the ``trunks'' of the magnetic trees? Will they just stay
there, ``bleeding''? Will they reconnect and be drawn back below the convective
zone? Or will they also decay? As indicated above, the turbulent erosion models
(\citeNP{Petrovay+FMI:erosion}) seem to support the latter possibility. At any
rate, thorough observational studies of the active region decay (cf.\ the
review by \citeNP{vDG:review}) may shed more light on this problem.

\bigskip
\bigskip

\szekcio{5. CONCLUSION: ...AND THE SIMULATIONS?\vspace{-0.5 cm}}
As we have seen, there is still no really convincing answer to the question
``What makes the Sun tick?''. Several conflicting approaches exist, none of
which can fully explain the observed features of the solar cycle and all of
which rely on some more or less arbitrary assumptions. In contrast to the
optimistic outlook of about three decades ago, nothing seems safe now, not even
the dynamo wave origin of the sunspot butterfly diagram. And the prospects do
not seem to promise a spectacular change in this situation in the near future.

A number of previous reviews (e.g.\ \citeNP{Weiss:CUPrev}) expressed the view
that the ultimate solution to the solar dynamo problem should be expected from
numerical hydrodynamical simulations. So the reader may ask: why was the
simulational approach hardly even mentioned in this review? Why is it that only
mean field models were treated? The answer does not lie in the (undeniable)
bias of the author but in the fact that in the past decade there simply have
not been any numerical simulations constructed with the ambition of producing a
global model for the solar cycle. As we indicated above, such attempts were
made back in the 1980's; but, after their failure in reproducing the butterfly
diagram and the internal rotation profile became clear, this line of research
was completely abandoned. Those very few hydrodynamical simulations in the
1990's that had some direct bearing on the solar dynamo problem aimed at the
modelling of smaller volumes inside the convective zone with the main purpose
of studying the fine structure and transport of the magnetic field
(\citeNP{Nordlund+:undershoot}, \citeNP{Tobias+:pumping},
\citeNP{Dorch+Nordlund:pumping}).

The belief in the ``miraculous healing power'' of simulations may indeed have
been too zealous before. After all, even in an ideal future where an infinite 
computing potential would make it feasible to make a 3D direct numerical
simulation of the whole Sun for $4.6\cdot 10^9$ years, the result would have no
more direct benefit for us (apart from demonstrating that the laws of classical
magnetohydrodynamics are indeed sufficient to describe solar phenomena) than
offering a chance to determine any physical quantity at any internal point of
our star with arbitrary precision. This feat, though far beyond the
possibilities of contemporary observational solar physics, is still basically
experimental work that, though important, in no way can substitute ``real''
theory. 

And yet, the other extreme, represented by the present complete abandonment of 
the simulational approach, is just as lamentable as its opposite. This is
especially so as the great advance in the geodynamo problem I referred to in
the Introduction can mainly be attributed to the success of numerical
simulations. The following pseudo-historical quote by Borges fits nicely the
present  situation in this respect.

\bigskip

\noindent
\fbox{
\begin{minipage}{\columnwidth}
\sl
Jorge Luis Borges:
\smallskip

\centerline{On Rigour in Science}

\bigskip

...In that Empire, the Art of Cartography achieved such Perfection that the
map of a single Province occupied a whole City, and the map of the Empire, a
whole Province. With time, these Gigantic Maps proved unsatisfactory and the
Colleges of Cartographers set up a Map of the Empire that had the size of
the Empire and coincided with it point by point. Less addicted to the Study of
Cartography, the Following Generations considered this extensive Map useless
and, not without Disrespect, they abandoned it to the Mercy of the Sun and of
the Seasons. In the Western deserts broken Ruins of the Map still persist,
inhabited by Animals and Beggars; in all the Country no other Relicts of the
Geographic Disciplines remain.\\
\strut\hfill {\small{\sc Su\'arez Miranda}: {\it Viajes de varones prudentes,}\\
\strut\hfill {\rm libro cuarto, cap.XLV, L\'erida, 1658}}

\end{minipage}

}

\rm

\szekciononu{Acknowledgements}
This work was funded by the FKFP project no.~0201/97, and the OTKA project 
no.~T032462.


\end{document}